\documentclass {article}
\usepackage{graphicx}

\begin{document}

\title{Compact anisotropic stars with membrane - a new class of exact solutions to the Einstein field equations}

\author{Michael Petri\thanks{email: mpetri@bfs.de} \\Bundesamt f\"{u}r Strahlenschutz (BfS), Salzgitter, Germany}

\date{{June 16, 2003 (v1)} \newline {May 1, 2004 (v3)}}

\maketitle

\begin{abstract}
I report on a new class of solutions to the classical field
equations of general relativity with zero cosmological constant.
The solutions describe a spherically symmetric, compact self
gravitating object with a smooth interior matter-distribution,
residing in a static electro-vacuum space time.

An outstanding feature of the new solutions is a sharp,
non-continuous boundary of the matter-distribution, which is
accompanied by a membrane consisting of pure tangential pressure
(surface tension/stress). The interior matter state generally has
a locally anisotropic pressure.

A procedure to generate various new solutions is given. Some
particular cases are derived explicitly and discussed briefly. To single
out the physically most promising solutions, a selection principle
based on the holographic principle is formulated.

One solution of particular interest emerges. The so called
holographic solution, short holostar, is characterized by the
property, that the "stress-energy content" of the holostar's
boundary membrane is equal to its gravitating mass. The holostar's
real membrane has identical properties to the - fictitious -
membrane attributed to a black hole by the membrane paradigm. This
feature guarantees, that the holostar's dynamical action on the
exterior space-time is practically identical to that of a black
hole with the same mass. The holostar's interior matter state can
be interpreted as a collection of tightly packed radial strings
attached to the spherical boundary membrane. The properties of the
holographic solution are discussed in detail in three parallel
papers. The solution provides a singularity free alternative to
black holes and the standard cosmological models.

\end{abstract}

\maketitle

\section{\label{sec:intro}Introduction}

Black holes are among the simplest objects of theoretical physics,
fully describable by just a few parameters. According to our
present knowledge, a black hole is the final, most compact state
of any sufficiently large gravitationally bound object.

Because of their subtle links to phenomena of quantum origin, such
as Hawking radiation and entropy, many researchers view black
holes as important theoretical probes to explore some aspects of
the "new" physics beyond the Standard Model. The study of black
holes plays a major role not only in classical relativity, but
also in loop quantum gravity (LQG) and string theory. Despite much
progress several open questions of black hole physics remain,
mostly relating to their inner structure.

Our limited understanding of self gravitating objects can be
traced to the fundamental problem, that so far there is no
satisfactory theoretical foundation from which the "matter side"
of gravity can be constructed by first principles.

A - possibly subjective - survey of recent and not so recent
research results seems to point to the direction, that the concept
of boundary areas might play a fundamental role in our future
understanding of gravity, on the classical as well as on the
quantum level.

Bekenstein was probably the first to realize the importance of
boundary areas in classical general relativity. His conjecture,
that the entropy of a black hole should be proportional to its
event horizon \cite{Bekenstein/72} was spectacularly confirmed by
Hawking \cite{Hawking/75}, who demonstrated that black holes have
a definite temperature, by which the constant of proportionality
could be fixed.

In the recent past insights of a similar kind were achieved on the
quantum level. The discovery of area-quantization in loop quantum
gravity \cite{Rovelli/Smolin} can be viewed as one of the most
important achievements in this field. In essence,
area-quantization in non-perturbative background-free LQG is
derived from the assumption, that the physical world is
fundamentally relational (no background- or pre-geometry), and
should most naturally be described in terms of quantities which
are invariant with respect to the artifacts of the particular
description that we impose. The relevance of boundary areas has
also become apparent in the string approach. Certain solutions of
string theory  suggest, that the dynamical degrees of freedom in
an internally consistent string theory are proportional to the
boundary of a space-time region. This conjecture has become known
to a larger audience under the popular name of "holographic
principle" \cite{Hooft/hol, Susskind/hol}.

Whereas the "geometric" side of gravity can be considered to be
fairly well understood, the incorporation of "matter" into gravity
is still a matter of debate. The question with respect to the
fundamental (matter) degrees of freedom in a self-consistent
theory of gravity has not been answered. The black hole solutions
of classical general relativity give paradoxical answers. Whereas
the fundamental degrees of freedom of a black hole are associated
with its horizon, the "mass" of the space-time is associated with
the central singularity. In string theory the fundamental degrees
of freedom are known, at least in principle. But we have not (yet)
been able to locate a single string (or brane) in the physical
world. The loop approach so far has focussed on the geometry,
which is quantized without reference to matter. Although there is
some evidence that the fundamental results of LQG, such as area
and volume quantization, will remain valid if matter is added to
the theory \cite{Rovelli/measurement, Smolin/measurement}, it is
not yet clear how this is to be done.

Loop quantum gravity gives beautiful answers to what the outcome
of the measurement of an area will be, but it leaves physicists
quite in the dark "what area" actually should be measured. Area
operators referring to different (coordinate) surfaces generally
do not commute \cite{Rovelli/primer}. Intuitively we are
accustomed to define the location of a surface via matter.
Diffeomorphism invariance however suggest, that any meaningful
area must refer to matter. Unfortunately the classical vacuum
black hole solutions of general relativity do not contain matter
(or fields) that could be associated with a surface.

The problem of "what we should measure" is a fundamental one. Many
researchers view the event horizon of a black hole as the surface,
to which some answers of quantum gravity, such as the quantization
of the area, should be applied.\footnote{see for example
\cite{Bekenstein/Mukhanov}} However, this appears not altogether
consistent and happens to be in direct contradiction to one of the
most successful fundamental principles of physics, relationism.
Relationism demands that space-time events, or more generally
space-time regions, must be defined by
matter.\footnote{Alternatively: by localizable properties of the
fields, which may be just the other side of the same coin.} The
famous "hole argument" of classical general relativity, given by
Einstein in 1912 \cite{Einstein/Grossmann} demonstrates quite
clearly, that the measurement of a distance, an area or volume in
empty space doesn't make sense, i.e. doesn't yield an unambiguous
result. If relationism is a fundamental property of the physical
world, and if areas play an important role in gravity, any
physically meaningful area should be defined by matter and should
be localizable through matter.

What form could such matter-states take? It has been known for
quite some time that in a spherically symmetric context different
space-times regions, such as an interior de-Sitter core and an
exterior electro-vac region, can be smoothly matched through a so
called transition region. The not necessarily thin transition
layer generally has a substantial surface pressure/tension, whose
integrated "energy-content" can become comparable to or even
exceed the asymptotic gravitational mass of the object (for a very
general discussion of such space-times, encompassing rotating
sources, see \cite{Burinskii} and the references therein). In
\cite{Giambo} formula for the surface energy-density and surface
pressures for an infinitesimally thin transition region were
given, however no explicit solutions were derived. One particular
solution, the so called "gravastar", that has been been put
forward recently by \cite{Mazur/Mottola}, has received some
considerable interest. The gravastar consists of a localizable
thin, spherical shell of matter separating an interior de Sitter
condensate phase from an exterior Schwarzschild vacuum. The
entropy of the gravastar, however, is calculated to be orders of
magnitude lower than the Hawking entropy of a black hole of the
same gravitating mass. Whereas the Hawking entropy scales with
area ($S \propto r^2 \propto M^2$), the entropy of the gravastar
scales with its square-root ($S \propto r \propto M$). It will
remain to be seen, whether such a substantial deviation from the
established theoretical framework of black hole physics will find
acceptance.

The concept of Hawking entropy is theoretically well founded and
has been derived through several independent ways. The essential
result, the entropy-area law, has its foundations in the
fundamental "Four Laws of Black Hole
Mechanics"\cite{Bardeen/Carter/Hawking, Bekenstein/72,
Bekenstein/73}. The entropy-area law has been confirmed by
calculations in string theory \cite{Strominger/Horowitz,
Maldacena/Strominger, Strominger/Vafa}, as well as in the loop
approach \cite{ABCK}. Therefore in this work the position is
taken, that the thermodynamic properties of a compact, self
gravitating object are - at least approximately - correctly
described by the Hawking-entropy and -temperature, irrespective of
its interior structure.

However, identifying the enormous entropy of a black hole with its
event horizon, a vacuum region locally indistinguishable from the
surrounding vacuum space-time, whose location can only be
determined by (global) knowledge of the whole space-time's future,
seems to be incompatible with our well established conceptions
about causality\footnote{The event horizon "moves" a-causally in
anticipation of the matter that will eventually pass it in the
future.} and the fundamental nature of entropy. In fact, the
statistical entropy of a macroscopic system is amongst the most
rigorously defined physical concepts, exactly calculable if the
microscopic constituents and their interactions are known.

This leaves us with several open questions. What is the true
origin of the Hawking-entropy of a black hole? Where do the
microstates of a black hole reside and how do they interact?
Where has the constituent matter gone? Does it and how does it
participate to the entropy and to the asymptotic gravitational
mass and/or field? What is the origin of (gravitational) mass?
Are the sources of the gravitational field localizable, and if
yes, to what extent (to a point, string or surface)? What role do
boundary areas play? And finally: What are the fundamental
"matter" states, from which the matter side of gravity can be
constructed from first principles?

Recent research in string theory, as well as in loop quantum
gravity, have provided some very important insights. Classical
general relativity so far has not been very successful in
providing satisfying and consistent answers. Classical general
relativity has been plagued with various (apparent) paradoxes,
most of which can be traced to presence of an event horizon in
vacuum and its "one-way membrane-property".

In this paper exact solutions to the field equations are derived,
which might help to resolve some of the paradoxes. The properties
of the solutions might also be helpful to advance the
understanding of the "matter side" of gravity.

\section{\label{sec:field_equations}Field Equations for a Spherically Symmetric System with Locally Anisotropic Pressure}

The approach taken in this paper is to derive the components of
the stress-energy tensor for the most general, isolated,
spherically symmetric self gravitating object. Neither will I
assume that the pressure is locally isotropic, nor that the
matter-fields (i.e. mass-density $\rho$ and the three principle
pressures $P_i$) are continuous. However, the metric is assumed to be continuous
everywhere.

With respect to notation I closely follow the presentation in
\cite{Weinberg/GR}, using the standard spherical metric in the (+ - -
-) sign-convention, with geometric units $ c = G = 1$:

\begin{equation}
ds^2 = B(r) dt^2 - A(r) dr^2 - r^2 d\theta^2 - r^2 \sin^2 \theta d\varphi^2
\end{equation}

The most general form of the stress-energy tensor for a
spherically symmetric (static) system is constructed out of three
independent matter/pressure fields, which, due to spherical
symmetry, do not depend on $ \theta $  and $ \varphi $, have the
same magnitude in $\partial \theta$-  and $\partial \varphi
$-direction, and, due to the requirement of staticness, are
independent from a suitably chosen - time coordinate t:

\begin{equation}
T_\mu^\nu(r) = \left( \begin{array} {cccc}
\rho(r) \\
& -P_r(r) \\
&& -P_\theta(r) \\
&&& -P_\theta(r) \\
\end{array} \right)
\end{equation}

Due to spherical symmetry the trace of the stress-energy tensor,
$T$, only depends on the radial coordinate $r$. The $r$-dependence
will be dropped in the formula, whenever appropriate:

\begin{equation}
T = T_\mu^\mu = \rho -P_r - 2 P_\theta
\end{equation}

With $T$ the field equations read as follows:

\begin{equation}
R_{\mu \nu} = -8 \pi \left( {T_{\mu \nu} - \frac {T} {2}}
\right)
\end{equation}

The components of the Ricci-tensor can be calculated from the
metric coefficients and their first and second derivatives. Due to
spherical symmetry the field equations reduce to three equations
for the three matter-fields:\footnote{see for example \cite[p.
300]{Weinberg/GR} or \cite[p. 128]{Flie/AR}}

\begin{equation} \label{eq:R00}
R_{tt}=R_{00}=-\frac{B''}{2A} + \frac{B'}{4A} \left(
{\frac{A'}{A} + \frac{B'}{B}} \right) - \frac{B'}{rA} = - 4 \pi B
(\rho + P_r + 2 P_\theta)
\end{equation}

\begin{equation} \label{eq:R11}
R_{rr}=R_{11}=\frac{B''}{2B} - \frac{B'}{4B} \left(
{\frac{A'}{A} + \frac{B'}{B}} \right) - \frac{A'}{rA} = - 4 \pi A
(\rho + P_r - 2 P_\theta)
\end{equation}

\begin{equation} \label{eq:R22}
R_{\theta\theta}=R_{22}=-1 - \frac{r}{2A} \left(
{\frac{A'}{A} - \frac{B'}{B}} \right) + \frac{1}{A} = - 4 \pi r^2
(\rho - P_r)
\end{equation}

\section{\label{sec:general}A general procedure for the generation of spherically symmetric solutions}

Combining equations (\ref{eq:R00}) and (\ref{eq:R11}) gives the
following expression:

\begin{equation} \label{eq:AB'/AB}
\frac{A'}{A} + \frac{B'}{B} = \frac{(AB)'}{AB}=(\ln(AB))'= 8 \pi
A r (\rho + P_r)
\end{equation}

Thus, in a spherically symmetric space-time, any extended region
with $AB = const$ gives rise to an equation of state $\rho + P_r
= 0$ and vice versa.

Equation (\ref{eq:AB'/AB}) allows us to eliminate $B'/ B $ in
equation (\ref{eq:R22}), giving the following differential
relation between $A$ and $\rho$:

\begin{equation} \label{eq:r/A}
(\frac{r}{A})' = \frac{1}{A} - \frac{rA'}{A^2} = 1- 8 \pi r^2 \rho
\end{equation}

It is quite a remarkable feature, due to spherical symmetry, that
the radial metric coefficient, $A$, only depends on the
mass-density, $\rho(r)$, even in the case of anisotropic pressure.

Integration of (\ref{eq:r/A}) gives the well known expression for
the radial metric coefficient of a spherically symmetric
gravitationally bound object:

\begin{equation} \label{eq:A}
A(r) =  \frac{1}{1 - \frac{2M(r)}{r}}
\end{equation}

with

\begin{equation} \label{eq:M}
M(r) = M_0 + \int_0^{r} {4 \pi r^2 \rho(r) dr}
\end{equation}

A point mass $M_0$ at the origin has been included as an
integration constant.

The time coefficient of the metric $B$ can be calculated from $A$
and the two matter fields $\rho$ and $P_r$ by means of integrating
equation (\ref{eq:AB'/AB}). The integration is usually performed
starting at $r=\infty$ and setting $B(\infty)=1$. From a
fundamental viewpoint it is more natural to take the event horizon
(when it exists) as the starting point for any integration of the
field equations.

The event horizon with its finite proper area and fixed topology
is an almost ideal "anchorpoint" for the space-time geometry, at
least for the spherically and axially symmetric vacuum solutions.
If there is vacuum outside the event horizon, the starting values
for the integration of equations (\ref{eq:r/A}) and
(\ref{eq:AB'/AB}) at the event horizon ($r=r_+$) are unambiguously
known: $r_+/A(r_+) = 0$ and $ln{AB(r_+)} = 0$.

In the case of a space-time with no event horizon, the boundary of
the matter-distribution, $r_h$, should be taken as starting point
of the integration of equations (\ref{eq:r/A}) and
(\ref{eq:AB'/AB}). We then still have $ln{AB(r_h)} = 0$, if the
exterior vacuum space-time is asymptotically flat. If there is no
event horizon the integration constant $r_h/A(r_h) = r_0$ can in
principle take on any positive value. The more compact a
self-gravitating object becomes, the smaller $r_0$ will be. The
most compact self-gravitating object without event-horizon is
expected to have $r_0 \approx r_{Pl}$.

The general procedure for determining the metric of a compact,
spherically symmetric object is:

(i) Determine $A$ by the following definite integral:

\begin{equation} \label{eq:r/A:r0}
\frac{r}{A} = r_0 + \int_{r_h}^r{1- 8 \pi r^2 \rho dr}
\end{equation}

(ii) Determine $B$ by:

\begin{equation} \label{eq:AB'/AB:r0}
\ln(AB) = \ln{AB(r_h)} + \int_{r_h}^r{8 \pi
A r (\rho + P_r) dr}
\end{equation}

Although it is preferable to integrate the field equations
starting out from the boundary of the compact object, in this
paper more often than not the equation for $A$ ´will be integrated
in the "usual" way, i.e. starting out from $r=0$, using a point
mass $M_0$ at the origin as an integration constant. This will
make it easier to compare the parameters of the solution, such as
the value of its point mass at the origin, with the well known
Schwarzschild solution.

$A$ and $B$ can be derived by integrals involving only $\rho$ and
$P_r$ without any explicit reference to the tangential pressure
$P_\theta$. This seems to imply that the metric only depends on
$\rho$ and $P_r$. A closer analysis shows that the dependencies
are more subtle. The integration constant of physical importance
is $M_0$ (or rather $r_0$). The global scale factor for $B$ has no
physical significance. However, any specific choice of $M_0$ (or
$r_0$) fixes the tangential pressure and vice versa, so - in fact
- the metric coefficients $A$ and $B$ depend unambiguously on all
three matter-pressure fields.

The tangential pressure $P_\theta$ can be calculated from the
metric, either from equations (\ref{eq:R00}) and (\ref{eq:R11}) or
by the following equation:

\begin{equation} \label{eq:Ptan:AB}
8 \pi P_\theta = \frac{1}{AB}(\frac{B''}{2}+
\frac{B'}{r})- 2 \pi(\rho + P_r)(\frac{rB'}{B}+2)
\end{equation}

In the general case, however, the tangential pressure is most
conveniently determined by the continuity equation, which in the
case of locally anisotropic pressure ($P_r \neq P_\theta$) is
given by:

\begin{equation} \label{eq:cont0}
\frac{1}{A}( {P_r}' + \frac{2 P_r}{r})+\frac{B'}{2AB}(P_r+\rho)-\frac{2P_\theta}{rA} =0
\end{equation}

As long as $1/A$ has no zeros (i.e. a space-time without event
horizon) the tangential pressure $P_\theta$ can be derived from
the continuity equation via:

\begin{equation} \label{eq:cont}
P_\theta = {P_r} + \frac{r {P_r}'}{2}+\frac{r B'}{B}(P_r+\rho)
\end{equation}

If the metric is known, all matter fields can be derived from the
metric coefficients via equations (\ref{eq:r/A}, \ref{eq:AB'/AB},
\ref{eq:Ptan:AB}).

The matter-density $\rho$ and the radial pressure $P_r$ depend
only on the first derivatives of the metric, whereas $P_\theta$
depends also on the second derivatives. This has important
consequences on the structure of the solutions. The
matter-fields of the new solutions generally contain
step-discontinuities ($\theta$-distributions) and non-regular
$\delta$-distributions. But only the tangential pressure may have a
$\delta$-distribution, otherwise the metric could not be
continuous. Continuity of the metric is essential for the
mathematical structure of general relativity, whereas the
"gravitational field", i.e. the metric derivatives, may contain
finite jumps without severe mathematical consequences.

In the following $\rho$ and $P_r$ are assumed to have at most
finite jumps. This guarantees, that the event horizon
(alternatively: the boundary of the matter-distribution) can be
chosen without any ambiguity as starting point for the
integration of the metric coefficients $A$ and $B$.

The general procedure to generate solutions for the spherically
symmetric case, which will be followed in this paper, is the
following:

\begin{itemize}

\item Guess a radial dependence for the mass-density $\rho(r)$.
The mass-density can be non-continuous, but should not contain any
$\delta$-distributions.

\item Calculate the radial metric coefficient $A(r)$ by
integrating equation (\ref{eq:r/A}) with a particular choice of
point mass at the origin $M_0$. Alternatively a particular choice
of $r_0 = r_h / A(r_h)$ at the boundary of the matter-distribution
is used.\footnote{If the boundary of the matter-distribution,
$r_h$, coincides with the position of the event horizon, i.e. $r_h
= r_+$, the integration constant $r_0$ must be zero.}

\item Determine the radial pressure $P_r(r)$ from $\rho(r)$ by
assuming a particular equation of state. This allows one to
calculate the time coefficient of the metric, $B(r)$ via
integration of equation (\ref{eq:AB'/AB}). If the
matter-distribution is bounded and the exterior space-time
asymptotically flat, the integration constant for $B$ is given by
$\ln{AB}(r_h) = 0$.

\item Finally determine the tangential pressure $P_\theta(r)$ from
the first and second derivatives of the metric (\ref{eq:Ptan:AB})
or by the continuity equation (\ref{eq:cont}).

\end{itemize}

A vast number of solutions can be constructed by this procedure.
Any well behaved function for the energy-density $\rho$ and for
the equation of state $P_r(\rho)$ will lead to a solution. Various
analytical or numerical solutions can be found in the
literature.\footnote{Algorithms similar or equivalent to
the above described procedure can be found in the recent
literature. These rather general approaches including anisotropic
pressure states appear to have been developed quite recently and
apparently independently by different authors, see for example
\cite{Burinskii, Dev/00, Dymnikova, Elizalde, Giambo, Herrera,
Mak, Salgado}.}

In this paper I will briefly discuss some solutions, which I found
interesting from a physical or mathematical perspective. The
choice is subjective. I present the main arguments that have led
to the particular choice of solution, the so called "holographic
solution" of section \ref{sec:holo}, which is discussed in greater
detail in \cite{petri/hol, petri/charge, petri/thermo}. A thorough
discussion of some of other new solutions which might be of
interest must be left to forthcoming papers.

\section{\label{sec:a=-1}Solutions with equation of state: $\rho + P_r = 0$}

Solutions with the above equation of state can be regarded as a
natural extension of the Schwarzschild and
Reissner-Nordstr\"{o}m solutions:

The Schwarzschild and Reissner-Nordstr\"{o}m solutions satisfy the
constraint $AB=1$ throughout the whole space-time.From the
(multiplicative) constraint on the metric, $AB=const$, the
(additive) constraint on the fields, $\rho + P_r =0$, follows via
equation (\ref{eq:AB'/AB}).

A number of physically interesting solutions appear to lie in this
restricted class of solutions, or are approximated by this class.
If we assume that the mass-energy density $\rho$ is positive, the
radial pressure will be negative for this type of solutions. This
shouldn't be a problem: An equation of state with a negative
pressure is well established in inflational cosmology. Recent
results on the large scale distribution of matter in the universe
indicate a "dark matter" component with overall negative pressure
$p \simeq -0.8 \rho$ (see for example \cite[p. 244]{Peacock}),
which - within the errors - comes close to the equation of state
$\rho + P_r = 0$. Furthermore, an equation of state with negative
pressure has been proposed in the not so recent past for the
Lorentz invariant vacuum state, see for example
\cite{Zel'dovich/72}. $\rho+P_r=0$ also is the equation of state
for a cosmic string, with $\partial r$ being the longitudinal
direction of the string.

Imposing the constraint $\rho + P_r = 0$ on the solutions greatly
simplifies the math. Equations (\ref{eq:R00}), (\ref{eq:R11}) and
(\ref{eq:R22}) are reduced to the following set of equations:

\begin{equation} \label{eq:rB'} {(rB)}' = 1 - 8 \pi r^2 \rho
\end{equation}

and

\begin{equation} \label{eq:Ptan:B} 8 \pi P_\theta = \frac{B''}{2} + \frac{B'}{r}
\end{equation}

By differentiating equation (\ref{eq:rB'}) or by using the
continuity equation (\ref{eq:cont}) the tangential pressure can
be expressed solely in terms of the mass-density:

\begin{equation} 8 \pi P_\theta = \frac{B''}{2} + \frac{B'}{r} =
\frac{{(rB)}''}{2r} = -8 \pi (\rho + \frac{r \rho'}{2})
\end{equation}

The system of three nonlinear equations in the two metric
coefficients $A$ and $B$ and their first and second derivatives has
been transformed into a linear, albeit inhomogeneous set of
differential equations of second order in $B$.

Equation (\ref{eq:rB'}) has the following - well known - solution:

\begin{equation}
B(r) = 1 - \frac{2 M(r)}{r}
\end{equation}

I will now make a rather special ansatz for the matter-density
$\rho(r)$:

\begin{equation} \label{eq:rho:1}
\rho=\frac{c}{8 \pi r^2} \theta(r-r_+)
\end{equation}

I.e. the mass-density follows an inverse square-law up the
horizon, $r_+$, where it drops to zero in one discontinuous step.
$c$ is a dimensionless constant, not to be confused with the
velocity of light. Note that in this section the boundary of the
mass-distribution, which will be denoted by $r_h$, coincides with
the radial coordinate position of the gravitational radius of the
object, $r_+ = 2M$. This is not mandatory. In the following
sections solutions with $r_h \neq r_+$ will be studied.

With hindsight, the ansatz in equation (\ref{eq:rho:1}) can be
"justified" by the following arguments:

\begin{itemize}

\item Avoidance of trapped surfaces

We would like to avoid trapped surfaces in our solutions. On the
other hand, we are interested in the most compact, bounded
matter-distribution possible. The second condition requires that
the exterior space-time is vacuum up to (or almost up to) the
gravitational radius of the object, which in the case of $r_+ =
r_h$ coincides with the event horizon. To fulfill the first
condition, i.e. to avoid trapped surfaces, the interior matter
must extend to the event horizon with $\rho \geq 1 / (8 \pi
{r_+}^2)$ at the event horizon. This can be seen as follows:

If there were vacuum at the "inner" side of the event horizon, the
coefficients $A$ and $B$ of the exterior Schwarzschild metric and
their derivatives would be continuous throughout the horizon. Due
to the finite slope of $1/A$ at the horizon and $1/A(r_+) = 0$ the
radial coordinate inevitably becomes time-like for $r < r_+$. A
time-like coordinate cannot stand still for any material object,
thus all radial distances must shrink. Vacuum "behind" the horizon
implies that any concentric sphere within this vacuum region is a
trapped surface. From the singularity theorems it follows that a
singularity inevitably will form. If trapped surfaces and
therefore singularities are to be avoided a sufficiently positive
mass-density must be "placed" directly at the inner side of the
horizon in such a way, that the jump in the mass density will
"turn up" the slope of $1/A$ such, that $1/A$ does not undergo a
sign change. It is easy to see from equation (\ref{eq:r/A}), that
any mass density $\rho \geq 1 / (8 \pi {r_+}^2)$ will do.

\item Scale invariant mass-density

The gravitational structures in the universe appear to be
scale-invariant, at least on scales large enough so that gravity
has become the dominant mechanism for structure formation.
Whenever possible, a scale-invariant mass-energy density should be
preferred in the solutions.

A mass-density following an inverse square law is scale invariant
(at least in a spherically symmetric context). In units $G=c=1$
mass (or energy) has dimensions of length. Thus from dimensional
considerations a "natural" mass-energy density is expected to have
dimensions of inverse length squared. Only such a mass-energy
density is truly scale-invariant.\footnote{This is not quite true:
In units $c=\hbar=1$ frequently used by particle physicists, mass
has dimensions of inverse length. Therefore an energy density
$\rho \propto 1/r^4$ can be considered as scale-invariant as well,
however only on a microscopic, not on a macroscopic scale.} Any
other functional form of the mass density requires at least one
dimensional constant, which would introduce a specific length
scale.

\end{itemize}

With $\rho$ given by equation (\ref{eq:rho:1}) the mass-function
$M(r)$ can be determined by a simple integration:

\begin{equation}
M(r)= \{ {\begin{array}{ll}
M_0 + c \frac{r}{2}   & r < r_+ \\
M_0 + c \frac{r_+}{2} & r \geq r_+
\end{array}}
\end{equation}

By another integration one arrives at the following expression for
the time coefficient of the metric, $B$:

\begin{equation} \label{eq:B:M0:r+}
B(r)= \{ {\begin{array}{ll}
1 - c - \frac{2 M_0}{r} & r < r_+ \\
1 - \frac{2 M_0 + c r_+}{r} & r \geq r_+
\end{array}}
\end{equation}

This result can be expressed by means of observable quantities,
such as the gravitating mass $M$ of the black hole or -
alternatively - its gravitational radius $r_+ = 2 M$:

Outside of the source-region, i.e. for $r > r_+$ the solution must
be identical to the well known Schwarzschild vacuum-solution, due
to Birkhoff's theorem. By comparing the Schwarzschild solution
with the above solution, one can relate the gravitating mass $M$
of the Schwarzschild solution to the parameters of the solution
given by equation (\ref{eq:B:M0:r+}):

\begin{equation}
M = M_0 + \frac{c r_+}{2}
\end{equation}

The well known identity between gravitational radius and
gravitating mass, $ r_+ = 2M $, then gives the following relation
between the "interior" parameter $M_0$ (point mass at the origin)
and the measurable "exterior" parameter $M$ (gravitating mass of
the black hole) and the dimensionless scaling factor, $c$, for the
mass-density:

\begin{equation} \label{eq:M0}
M_0 = M (1-c)
\end{equation}

The case $c=0$ is special. Here the point mass $M_0$ at the origin
exactly equals the gravitating mass $M$ of the black hole. The
interior matter-fields are zero. This is what is expected, if the
$c=0$ solution corresponds to the known Schwarzschild vacuum
solution.

In the following the metric will be expressed in terms of $r_+$
and $c$:

\begin{equation}
B(r)= { \{ {\begin{array}{ll}
(1-c)(1- \frac{r_+}{r}) & r < r_+ \\
1- \frac{r_+}{r} & r \geq r_+
\end{array}} }
\end{equation}

The tangential pressure can be derived from the metric. For
calculational purposes it is convenient to combine the interior
and exterior solutions with help of the Heavyside-step function
($\theta$-distribution) into one equation:

\begin{equation}
B(r)= (1-\frac{r_+}{r})(1-c+c\theta(r-r_+))
\end{equation}

With equation (\ref{eq:Ptan:B}) and by exploiting the well known
properties of the $\delta$- and $\theta$-distributions, i.e.
$\theta' = \delta$ and $B(r_+) \delta(r-r_+) = 0$, we arrive at
the following result for the tangential pressure:

\begin{equation}
P_\theta = \frac{c}{16 \pi r_+} \delta(r-r_+) = \frac{M-M_0}{32
\pi M^2} \delta(r-r_+)
\end{equation}

The reader can easily check by substituting $B(r)$, $A(r)=1/B(r)$,
$\rho(r)$, $P_r(r)$ and $P_\theta(r)$ into the field equations
(\ref{eq:R00}), (\ref{eq:R11}) and (\ref{eq:R22}), that the just
derived solutions satisfy the original field-equations and the
continuity equation exactly in a distributional sense.

The peculiar behavior of the tangential pressure is one of the
outstanding features of the new solutions, e.g. $c \neq 0$. An
inverse square law for the interior energy-density $\rho$ leads to
an interior tangential pressure identical zero, except for a
$\delta$-distribution at the horizon. The horizon, constituting
the boundary of the matter-distribution, is localizable, either
through its property as the dividing line between the
matter-filled interior region and the exterior Schwarzschild
vacuum region, or alternatively by the membrane's surface
tension/pressure. The interior matter is subject to a negative,
purely radial pressure, identical to the pressure of a classical
string in the radial direction ($P_r = -\rho; P_\perp = 0$).

The active gravitational mass of the interior space-time is zero,
as expected from the equation of state of a cosmic string. Thus it
appears as if the "net energy content" of the interior
matter-fields comes out zero. However, this does not take into
account the "energy content" of the horizon and the point mass
$M_0$ at the center.

The value of $c$ determines the tangential pressure at the
position of the horizon. The integration constant $M_0$ depends
directly on $c$ and on the gravitating mass $M$ via equation
(\ref{eq:M0}). Therefore $M_0$ cannot be chosen independently from
the tangential pressure. In fact, the tangential pressure at the
space time's boundary and the point mass at the origin are
intimately related. From a physical point of view it is the
tangential pressure at the event horizon which determines $M_0$.
The important physical question therefore is, what value the
tangential pressure should have at the horizon (or alternatively,
what value the point mass $M_0$ at the center should have.)

This question can only be answered in an affirmative way by
comparing the properties of solutions with different $c$, i.e.
different "horizon-pressure" (or different point-masses at the
origin), and choosing the solution whose properties are a better
approximation to the observation.

Yet theory might give some guidance in the selection of the
physically most relevant solution. There are two solutions, $c=0$
and $c=1$, which "stand out".

The "historic" Schwarzschild solution ($c=0$) has been intensively
studied. Its "horizon pressure" is zero, giving the
Schwarzschild-solution the advantage of an analytic metric, except
for the the singular point at $r=0$. The Schwarzschild solution
has the severe disadvantage of a singularity at $r=0$ (breakdown
of predictability). But both phenomena are related: In a
spherically symmetric space-time with event horizon and with an
equation of state $\rho + P_r=0$ a zero horizon pressure ($c=0$)
is necessarily accompanied by a large point-mass at the origin
($M_0 = M$).

For the $c=1$ solution the event horizon consists of a real
physical membrane with tangential pressure $P_\theta = 1 / (16 \pi
r_+)$. Due to the non-zero "horizon pressure" the metric is not
analytic at the horizon, not even differentially continuous.
However, there is no point-mass at the origin and thus no
singularity, which makes this type of solution particularly
attractive from a physical perspective.

Quite remarkably, the pressure of the membrane for the $c=1$ case
is exactly equal to the tangential pressure attributed to a black
hole by the membrane paradigm \cite{Thorne/mem}. According to the
membrane paradigm for black holes all (exterior) properties of a
(vacuum) black hole can be explained by a (fictitious) membrane
situated at the event horizon. Therefore, viewed by an exterior
observer, both situations, $c=0$ and $c=1$ are completely
indistinguishable. From a physical point of view it appears very
attractive to substitute the fictitious membrane with the real
membrane of the $c=1$ solution. This leaves all important results
derived for black holes with respect to the exterior space-time
unchanged, and at the same time introduces exactly what appears to
be needed (a zero point mass at the origin) to get rid of the
severe problems regarding the interior space-time of a classical
black hole, which is characterized by singularities, trapped
surfaces and causally-disconnected regions. However, we will see
in section \ref{sec:holo} that getting rid of space-time
singularities is not quite as easy as it seems. It requires one
other ingredient: The elimination of the event horizon.

\subsection{\label{sec:superpos}The weighted superposition principle}

The equation of state $\rho+P_r=0$, which is equivalent to $AB =
const$, allowed us to linearize the Einstein-equations in the
spherically symmetric case.

The linearity of the equations allows us to generate new solutions
by (weighted) superposition. If $B_1$ is a solution for the fields
$\rho_1, {P_r}_1$ and ${P_\theta}_1$ and $B_2$ is a solution for
the fields $\rho_2, {P_r}_2$ and ${P_\theta}_2$ , then the
weighted sum of the metric coefficients $c_1 B_1 + c_2 B_2$ will
be a solution for the respective weighted sum of the fields $c_1
\rho_1 + c_2 \rho_2, c_1 {P_r}_1 + c_2 {P_r}_2$ and $c_1
{P_\theta}_1 + c_2 {P_\theta}_2$, if the "norm condition" $c_1 +
c_2 = 1$ is met. It should be noted, that even though $B_1$ and
$B_2$ are genuine solutions of the field equations, $c_1 B_1$ or
$c_2 B_2$ generally are not, because the field equations are
inhomogeneous in the absence of sources. However, for
the "fundamental" mass-density $\rho = 1 / (8 \pi r^2)$ the
equations are rendered homogeneous.

Before some more general solutions to the equation of
state $\rho + P_r = 0$ are derived, I will briefly discuss three special
cases.

\subsection {$c=0$: the Schwarzschild-solution}

For $c=0$ the mass-density and pressures are identical zero,
except at the center which contains a point mass equal to the
gravitating mass of the black hole. The metric is identical to the
well known Schwarzschild vacuum solution. Therefore the
Schwarzschild vacuum solution constitutes a special case within
the broader class of solutions characterized by $AB = const$ or
$\rho + P_r = 0$.

\subsection {$c=2$: the negative-mass solution}

This solution has a negative point mass at the origin, exactly
equal but opposite in sign to the gravitating mass, i.e $M_0 =
-M$. The interior metric coefficients $A$ and $B$ are the absolute
values of the respective metric coefficients of the classical
Schwarzschild vacuum solution.\footnote{The "negative mass"
solution can be derived by the weighted superposition principle of
section \ref{sec:superpos}: Superposition of twice the singular
metric solution ($M_0 = 0$), discussed in the next chapter, minus
the Schwarzschild solution ($M_0 = M$) gives the "negative mass"
solution. The weights, $2$ and $-1$, add up to $1$.} The radial
distance coordinate $r$ remains space-like and the time-coordinate
$t$ time-like throughout the whole space-time. However, there is
an ambiguity at the event horizon $r_+$, as $B(r_+) = 0$.

The weak, strong, dominant and null energy conditions are
satisfied at any space-time \textsl{point}, except at the origin
and at the horizon. The weak energy condition is only violated at
a single point, the origin. Disregarding this singular point one
could view the $c=2$ solution as physically acceptable. However,
in this work I take the position, that it doesn't make sense to
attribute meaning to any physical quantity evaluated at a
space-time point. Any physically meaningful space-time region will
have a boundary area greater than the minimum non-zero area
eigenvalue of loop quantum gravity. A violation of a physical
principle, such as the weak or positive energy condition at a
space-time point isn't considered problematic, as long as
the violation is confined to a region smaller than Planck size.
Turning the argument around, I propose that any physically
acceptable classical space-time should be subject to the following
conditions:

\begin{itemize}
\item In a physically acceptable classical space-time any physical
quantity must be evaluated with respect to a physically meaningful
space-time region.\footnote{Note that this condition implies, that
it must always be possible to associate a physical quantity with a
"physically meaningful" space-time region. In the classical
context the association will be through the proper integral of the
quantity over the region. Other types of association, such as the
improper integral, are possible. In the context of quantum gravity
the appropriate operators have to be used. Throughout this paper I
consider a "meaningful" space-time region to be a region of
roughly Planck-volume bounded by a surface of roughly Planck
area.}

\item Each relevant physical quantity or physical condition,
evaluated for any physically meaningful space-time region, must
lie within the accepted range for this quantity.\footnote{For
example, in a classical context the weak energy condition must
hold for any arbitrary region of the space-time larger than
Planck-size.}
\end{itemize}

With the above proposition in mind it is instructive to evaluate
some of the properties of the $c=2$ solution. The improper
integral of the mass-density, including the negative point-mass at
the origin, over any concentric interior sphere with "radius" $r$
is given by:

$$M(r) = -\frac{r_+}{2} + r$$

We find that the mass-function, a measure for the gravitational
mass of the region, is negative for any sphere with $r < r_+/2$.
The positivity of the energy is violated in this region.

The weak energy condition should be evaluated by the improper
integral:

$$\int{(\rho -P_r) 4 \pi r^2 dr} \geq 0$$

We find that the integrated weak energy condition is violated for
any interior space-time region including the origin, due to the
negative point mass situated at $r=0$.

Therefore with respect to the above proposition the $c=2$ solution
is not physically acceptable, although it satisfies the relevant
physical conditions at all of its space-time points, except at a
collection of space-time points of measure zero (here: the origin,
the horizon).

An analysis of the geodesic motion of a test-particle gives
further evidence, that the region $r < r_+/2$ is physically
unacceptable.

For the $c=2$ solution the time coefficient $B$ of the metric is
equal to the absolute value of the time coefficient of the
classical Schwarzschild interior metric. Thus the effective
potential given in equation (\ref{eq:Veff:beta}) of the Appendix
reads:

\begin{equation} \label{eq:Veff:a=-1} V_{eff} = |
 \frac{r - r_+}{r_i - r_+}|\frac{r_i}{r}\left(1 -
\beta_i^2 (1-\frac{r_i^2}{r^2})\right)
\end{equation}

$r_i$ is the position of the interior turning point of the motion
and $\beta_i$ is the tangential velocity of the particle at the
turning point of the motion, expressed as a fraction to the local
velocity of light.

Independent of the constants of the motion, $r_i$ and $\beta_i$,
the potential energy attains its minimum value at the horizon.
Particles close to the horizon will undergo radially bounded
oscillations around the radial position of the horizon, $r_+$. For
particles with an interior turning point $r_i$ of the motion close
to the horizon the oscillations are bounded, irrespective of
$\beta_i$, i.e. even for photons. Therefore we can expect, that
small (radial) density fluctuations in the vicinity of the horizon
will not be able to destabilize the black hole.

Any particle whose interior turning point $r_i$ of the
motion is less than $r_+/2$, has - formally - a radial velocity at
infinity, which is greater than zero, as can be seen by
evaluating the expression $\beta_r^2(r) = 1 - V_{eff}(r) \geq 0$
at $r = \infty$:

\begin{equation} \label{eq:ri:c=2}
r_i \leq \frac{r_+}{2-\beta_i^2}
\end{equation}

Therefore any particle emanating from $r < r_+/2$ can in principle
escape to infinity, although in general it will have to make it
over the angular momentum barrier situated in the exterior
space-time.

Particles with high $\beta_i$, i.e. with high tangential velocity
at the turning point of the motion, are "unbound" for radial
coordinate values larger than $r_+/2$. For photons with $\beta_i^2
= 1$ the inequality of equation (\ref{eq:ri:c=2}) becomes $r_i <
r_+$. Therefore any zero-rest mass particle, wherever situated in
the black hole's interior, can tunnel through the angular momentum
barrier and permanently escape. Most massless particles don't even
need to tunnel. They just pass over the barrier: For zero
rest-mass particles the angular momentum barrier is situated at $3
r_+/2$. The effective potential has a local maximum at this point,
which is given by $V_{eff}(3 r_+/2) = 4 / 27 (r_i/r_+)^2 /
(r_+/r_i-1)$. The effective potential has been normalized to
$V_{eff}(r_i) = 1$. Therefore whenever $V_{eff}(3 r_+/2) \leq 1$,
the photons need not tunnel through the angular momentum barrier,
because the radial component of their velocity at the barrier is
greater than zero. This is the case for all zero rest-mass
particles with:

\begin{equation}
r_i \leq
\frac{3}{2}\frac{(1+\sqrt{2})^{\frac{2}{3}}-1}{(1-\sqrt{2})^{\frac{1}{3}}}
r_+ \cong 0.894 r_+
\end{equation}

For highly relativistic non-zero rest-mass particles a similar
relation holds. The black hole can be expected to loose much of
its inner mass. On the other hand, the time to pass through the
horizon (either way) is infinite, as viewed by an asymptotic
exterior observer. Therefore classically this type of black hole
might still be considered stable.

Quantum mechanically the position of the horizon should be subject
to fluctuations on the order of the Planck length. The exterior
time $t$ to approach the horizon up to a Planck-length (starting
out from the angular momentum barrier at $r = 3r_+/2$)\footnote{At
and beyond the angular momentum barrier the time-dilation due to
the gravitational field is negligible, compared to the time
dilation at the horizon: $B(3r_+/2) = 1/3 \gg B(r_+ + r_{Pl})
\simeq r_{Pl}/r_+$.} is roughly given by $t \approx r_+
\ln{r_+/r_{Pl}}$. For a black hole of the size of the sun ($r
\approx 10^{38} r_{Pl}$) this time is less than one micro-second.

Therefore it can be expected, that the black hole will quickly
"loose" all of its "inner" particles.

This result must be interpreted with some reserve. There are two
main objections. First, loop quantum gravity, with its prediction
of quantized geometry, might not turn out to be the true quantum
theory of gravity. This position is particularly advocated by
string theorists. In this case there might not be any Planck-scale
fluctuations of the geometry near the horizon, so that particles
cannot "slip out" under the event horizon with $t \approx r_+
\ln{r_+/r_{Pl}}$. Second, in a space-time with pressure particles
generally don't move on geodesics, as the particles are subject to
non-vanishing pressure forces in their respective local inertial
frames.

The first objection is as yet undecided. With respect to the
second objection one might argue in a very handwaving manner, that
although the interior particles won't move on geodesics, it seems
quite unlikely that an interior particle acquiring a non-zero
velocity at for example $r_i \cong r_+/2$ will be slowed down
enough by the pressure to prevent its escape: For sufficiently
large black holes, both mass-density and pressure in the region
$r_+/2 \leq r \leq r_+$ become arbitrarily low. Therefore one
might expect that the pressure not very much reduces the
likeliness of escape. A detailed analysis will prove, whether this
handwaving argument is essentially correct.

\subsection {\label{sec:singularmetric}$c=1$: the "singular metric" solution}

An interesting case, at least mathematically, is the solution with
$c=1$. It has the following, well behaved fields:

\begin{equation} \label{eq:c=1:rho}
\rho = -P_r = \frac{1}{8 \pi r^2} (1-\theta(r-r_+))
\end{equation}

\begin{equation} \label{eq:c=1:Pt}
P_\theta = \frac{1}{16 \pi r_+} \delta(r-r_+)
\end{equation}

However, its metric is singular within the whole interior region:

\begin{equation} \label{eq:B:c=1:1}
B(r)= (1- \frac{r_+}{r}) \theta(r-r_+)
\end{equation}

$B(r)$ is identical zero within the interior according to equation
(\ref{eq:B:c=1:1}). $A(r)$ is given by the inverse of $B(r)$. Thus
within the entire black hole's interior "time stands still"
($B=0$) and all proper radial distances are infinite ($A =
\infty$).

It can be objected, that the above solution is not a genuine
solution to the field equations at all, but rather a limiting
case, separating the two genuine solutions classes with a negative
or a positive point mass $M_0$ at the center of symmetry
respectively. On the other hand, if the matter fields are regarded
as distributions, operating on the function space of (continuous)
test functions spanned by the metric coefficients, the solution
should be at least mathematically well defined.

Physically the above solution is not very satisfactory, due to the
singular metric.\footnote{Note however, that some of the
undesirable features of the solution might be mended. For example,
the bad behavior of the interior metric can be removed if the
signature of the metric is changed from Lorentzian to Euclidian at
the horizon. For a metric which is completely Euclidian (interior
and exterior), the field equations yield a non-singular solution
for the interior metric coefficients $A$ and $B$ with the
matter-fields given in equations (\ref{eq:c=1:rho},
\ref{eq:c=1:Pt}). The utterly "frozen" state of the interior
matter, due to $B=0$, might even be regarded as a desirable
feature of the solution. Some researchers have postulated a
"frozen state", a so called "Planck-solid", inside a black hole
\cite{PlanckSolid}. Other researchers have pointed out, that
quantum gravity predicts extended space-time regions with $B=0$
(see for example \cite[p. 92]{Davies}). Also note, that with "time
standing still" and "radial distances being infinite" quantum
mechanical vacuum fluctuations should be heavily suppressed, even
if the metric is allowed to fluctuate. This could provide an
snbr-explanation for the cosmological constant problem.} It is
doubtful, whether the "singular metric solution" can be viewed as
an approximate description of a real physical system.

\subsection {\label{sec:holo}The holographic solution}

A modification of the above singular solution with similar
properties, but a non-singular interior metric can be
constructed as follows:

As before let us consider space-time to be divided into two
regions, separated by a spherical boundary at radial coordinate
position $r_h$. The interior region is filled with "matter", the
exterior region is vacuum.

The mass density within the interior region is assumed to be
equal to the "natural" mass density $\rho = 1 / (8 \pi r^2)$ of
the "singular metric" solution:

\begin{equation} \rho(r)= \frac{1}{8 \pi r^2} (1-\theta(r-r_h))
\end{equation}

In contrast to the "singular metric" solution, which does not have
a point mass at the origin, a (small) point mass $M_0$ is assumed
at the center.\footnote{This is always possible, despite the
formula of equation (\ref{eq:M0}): $M_0 = (1-c) M$. This formula
was derived under the assumption that the boundary of the
object coincides with the event horizon. Here we assume,
that $r_h$ may lie at a different position as the gravitational
radius $r_+$, presumably outside of $r_+$.} Integration of
equation (\ref{eq:rB'}) gives the following interior solution:

\begin{equation} B_i(r)= \frac{-2M_0}{r}
\end{equation}

The exterior solution outside the matter-filled region must be
equal to the Schwarzschild solution, with a gravitating mass $M =
r_+ / 2$:

\begin{equation} B_e(r)= 1-\frac{r_+}{r}
\end{equation}

The metric must be continuous at the boundary $r_h$. The interior
and exterior metric coefficient $B$ shouldn't have a sign-change,
i.e. should be positive everywhere. If the point mass at the
origin $M_0$ is negative one can match the interior solution with
$B>0$ to the exterior solution, so that $r_h > r_+$, i.e. the
boundary lies outside of the object's gravitational radius $r_+$:

\begin{equation} r_h = r_+ + (-2M_0)
\end{equation}

It is not possible to determine $M_0$ in the context of classical
general relativity. With the two constants of the theory, $c$ and
$G$, no universal constant with the dimensions of mass (or length)
can be constructed. Therefore, in the purely classical context it
would be reasonable to assume that $M_0$ depends linearly on $M$.
This puts $r_h$ at a constant proportion with respect to $r_+$,
leading to a "scale" invariant black hole, which "looks" the same
on any scale. A natural choice would be $M = M_0$, which leads to
$r_h = r_+$, i.e. the classical Schwarzschild vacuum solution.

If $M_0$ is postulated to be independent of $M$, the only
"natural" classical choice - without a universal constant of the
dimension of mass - is $M_0 = 0$, which leads to the "singular
metric solution". Quantum theory, however, provides a third
universal constant of nature, $\hbar$. From the three constants
$\hbar$, $c$ and $G$ a universal constant with the dimension of
mass can be constructed. Therefore, if $M_0$ is postulated to be
constant and non-zero, the only natural choice that remains is the
Planck mass. For the following I will set

$$ -2M_0 = r_0$$

and assume, that $r_0$ is a positive quantity corresponding to the
Planck-distance, except for a numerical factor of order unity. The
tangential pressure then is given by:

\begin{equation}
P_\theta = \frac{1}{16 \pi r_h} \delta(r-r_h)
\end{equation}

with

$$r_h = r_+ + r_0 = 2 M + r_0$$

The holographic solution satisfies all of the field equations,
including the continuity equation. The exterior metric in the
matter-free vacuum region is exactly that of a black hole with a
gravitating mass $M = r_+ / 2$.

In the following pages the term compactar (= compact star) will be
used to distinguish between the two classes of self gravitating
systems: (i) solutions with event horizon (black holes) and (ii)
the new type solutions, which have their gravitational radius
marginally inside the matter-distribution, i.e. $r_h - r_+ = r_0
\approx r_{Pl}$.

The holographic solution appears to be the most interesting case
of a compactar from a physical perspective. Its properties are
discussed briefly in this paper, and somewhat more in detail in
\cite{petri/hol, petri/charge, petri/thermo}. It turned out, that
the holographic solution can explain many of the phenomena
attributed to black holes, such as the Hawking entropy and
temperature, and at the same time has much in common with the
universe as we see it today. Its obvious string-character suggests
that black holes, and possibly even the universe itself might be
constructed hierarchically out of its basic building blocks, which
appear to be strings and membranes (and particles). It therefore
seemed appropriate to give the solution a name by which it can be
referenced without much overhead. Henceforth I will refer to the
holographic solution as "holostar" (= holographic star).

\subsection {\label{sec:rho^n:a=-1:blackhole}Black hole solutions with equation of state $P_r = - \rho$ and a power law in $\rho$}

The emphasis in this section is on black hole solutions, i.e.
solutions for which the radial coordinate position of the boundary
$r_h$ and the gravitational radius $r_+$ coincide. The case $r_h
\neq r_+$ will be covered in the following section.

The mass-density $\rho$ is expressed as a power law in the radial
coordinate $r$:

\begin{equation} \label{eq:rho:n}
\rho = -P_r = \frac{c}{8 \pi r^2}(\frac{r_+}{r})^n (1-\theta(r -
r_h))
\end{equation}

In the following calculations the argument ($r - r_h$) in the
$\delta$- and $\theta$-distributions will be omitted.

The radial metric coefficient $1/A$ can be derived from the
mass-density (\ref{eq:rho:n}) by integrating equation
(\ref{eq:r/A})

\begin{equation}\label{eq:B:n}
B(r) = \frac{1}{A(r)}=\big(1-
\frac{r_+}{r}(1-\frac{c}{1-n})-\frac{c}{1-n}(\frac{r_+}{r})^n
\big)(1-\theta) + (1-\frac{r_+}{r})\theta
\end{equation}

This solution is valid for all powers $n$, with the exception
$n=1$, which is discussed separately. The solutions have been
constructed such that $B$ is continuous at the horizon. The
integration was started out from the horizon, setting the
integration constant $r_0 = 0$.

In general the solutions behave like a point-mass at the origin
with $M_0$ given by:

\begin{equation} \label{eq:M0:n}
M_0 = \frac{r_+}{2}(1- \frac{c}{1-n}) =M(1- \frac{c}{1-n})
\end{equation}

If $M_0$ is to be zero or close to zero for arbitrary $M$, this
requires $c \simeq 1-n$.

The tangential pressure can be derived from the metric coefficient
$B$ and its derivatives via equation (\ref{eq:Ptan:B}):

\begin{equation}\label{eq:Ptan:n}
P_\theta = \frac{c n}{16 \pi r^2}(\frac{r_+}{r})^n (1-\theta) + \frac{c}{16 \pi r_+}\delta
\end{equation}

For $n=1$ the metric and the tangential pressure are given by:

\begin{equation}
B(r) = \frac{1}{A(r)}=(1-\frac{r_+}{r}(1 - c\ln{\frac{r_+}{r}}))(1-\theta) +
(1-\frac{r_+}{r})\theta
\end{equation}

\begin{equation}
P_\theta = \frac{c r_+}{16 \pi r^3}(1-\theta) + \frac{c}{16 \pi r_+}\delta
\end{equation}

By combining the above solutions with the help of the weighted
superposition principle discussed in section \ref{sec:superpos}, many interior
solutions for a black hole with given gravitational mass $M$ can
be constructed. One can start out with any power in the
mass-density $\rho = c_i / (8 \pi r^{(2+i)})$ and get a valid
solution which obeys the equation of state $\rho + P_r = 0$. For
any mass-density $\rho$ which has a power-expansion in the radial
coordinate $r$, a solution obeying the above equation of state can
then be constructed by the weighted superposition principle of section
\ref{sec:a=-1}.

Not all of these solutions will be realized physically. At the
current state of knowledge, i.e. lacking a detailed account of the
physical and mathematical properties of the new solutions, it will
be quite impossible to give precise rules for the selection of the
physically interesting solutions. The holographic principle might
provide a valuable guideline. According to the holographic
principle, first formulated by t'Hooft in \cite{Hooft/hol} and
extended by Susskind \cite{Susskind/hol}, a three-dimensional
gravitational system should be describable solely in terms of the
properties of its two-dimensional boundary.

The following list of requirements should be viewed as a first attempt to
formulate selection principles for the physically interesting
solutions of classical general relativity along the lines of the
holographic principle:

\begin{itemize}
\item weak holographic selection principle:

All properties of the classical space-time which are measurable by an
external observer, such as gravitational mass or charge, should be
explainable solely in terms of the properties of the boundary
surface of a self gravitating object.

Therefore the "stress-energy content" of the membrane situated at the
boundary should be equal to the gravitating mass of the object.

\item strong holographic selection principle

A somewhat stronger requirement is the following: The sum of all
matter-fields of a self-gravitating object, i.e. $\rho + P_r + 2
P_\theta$, i.e. its active gravitational energy-density, should be
zero everywhere, except at the boundary. Furthermore the trace of
the stress-energy tensor of all fields, integrated over the full
space-time, should be equal to the gravitating mass of the object.

\end{itemize}

For the above derived solutions the "weak holographic selection
principle" requires that the tangential pressure be of the
following form:

\begin{equation}
P_\theta= f(r)(1-\theta) + \frac{1}{16 \pi r_+}\delta
\end{equation}

This leads to the condition $c = 1$, which is satisfied for a wide
class of solutions and thus is too weak to restrict the solutions
to a manageable number. If we require the point mass at the origin
to be zero, however, we get $n=0$ from equation (\ref{eq:M0:n}),
which gives us the singular metric solution.

In order to apply the "strong holographic selection principle" the
sum of all matter fields, i.e. the active gravitational
energy-density, is evaluated:

\begin{equation}
\rho + P_r + 2P_\theta = 2P_\theta = \frac{c n}{8 \pi
r^2}(\frac{r_+}{r})^n (1-\theta) + \frac{c}{8 \pi r_+}\delta
\end{equation}

If this sum is to be zero everywhere except at the boundary, the
condition $n=0$ follows.\footnote{This remains true for arbitrary
$\rho$, whenever $\rho$ can be expressed as a power-expansion in
$r$, because the basis functions $r^n$ are linearly independent}

Thus by invoking the strong and weak holographic selection
principles formulated above, only one solution ($c=1$, $n=0$)
remains. The "singular metric solution" of chapter
\ref{sec:singularmetric} has been recovered.

Note that the "singular metric" solution can also be singled out
by formulating constraints with respect to the $00$-component of
the Ricci-tensor:

$$ R_{00} = - 4 \pi B (\rho + P_r + 2 P_\theta) $$

The $R_{00}$-component of the Ricci-Tensor is identical zero
throughout the \textit{whole} space-time for the "singular metric"
solution (the delta-distribution is cancelled by the zero in
$B$). Thus demanding that $R_{00}$ be zero everywhere leads to the
condition $n=0$, however doesn't fix the value of $c$.

In the more general case ($\rho \neq c/(8 \pi r^2)$; $rho+P_r \neq
0$) the interior pressure is locally anisotropic. Generally, at
any position $r \neq 0$ the radial pressure component will differ
from the two tangential pressure components. In the case $P_\theta
\neq 0$ the following difficulty arises: At $r=0$ both $P_r$ and
$P_\theta$ can point in any direction, thus $P_\theta$ must be
considered as radial pressure component. If $P_r$ and $P_\theta$
have different both non-zero values at $r=0$, there are two
mutually incompatible values for the pressure at this particular
point. In a pure classical context this would lead to the
rejection of all solutions except for those, where one of the
pressure components is zero or both are equal at $r=0$. However,
the failure of classical concepts at a single space-time point
should not be viewed as an unsurmountable problem. According to
loop quantum gravity any region of space-time with a (closed)
boundary surface smaller than roughly the Planck-surface should be
viewed as devoid of interior structure and thus inaccessible for
measurement. Geometric operators referring to a point are
undefined in quantum gravity \cite{Smolin/advances}.

\subsection {\label{sec:rho^n:a=-1:compactar}Compactar solutions with equation of state $\rho + P_r = 0$ and a power law in $\rho$}

In this section I extend the holographic solution of section
\ref{sec:holo} to mass-densities which follow an arbitrary power
law in the radial coordinate $r$. The only difference to the
previous section is, that $r_h \neq r_+$.

With the mass density given by equation (\ref{eq:rho:n}) $B$ is
derived by a simple integration:

\begin{equation}
B = \big( 1 - \frac{2M_0}{r} - \frac{c}{1-n}{(\frac{r_+}{r})}^n
\big)(1-\theta) + \big(1 - \frac{r_+}{r}\big) \theta
\end{equation}

The metric must be continuous. Therefore we can determine the
position of the boundary $r_h$ as the position, where interior and
exterior metrics are equal:

\begin{equation}
r_h = r_+ {\big( \frac{1-n}{c}
(1-\frac{2M_0}{r_+})\big)}^{\frac{1}{1-n}}
\end{equation}

Any exponent $n > 1$ will either require a negative value of $c$,
i.e. a negative energy-density $\rho$ or a point mass $M_0 > r_+$,
in order to get a real valued solution.

Under the assumption that $M_0$ is universal and small and $r_+$
generally is large, $M_0$ must be negative and the coefficient
$\frac{1-n}{c}$ must be positive and greater or equal to one, if
the boundary $r_h$ of the matter distribution shall lie outside of
its gravitational radius $r_+$.

With the continuity equation (\ref{eq:cont}) the tangential
pressure can be calculated:

\begin{equation}
P_\theta = \frac{c n}{16 \pi r^2} {(\frac{r_+}{r})}^n (1-\theta) +
\frac{c}{16 \pi r_h} {(\frac{r_+}{r_h})}^n \delta
\end{equation}

It is possible to replace the term in front of the
delta-distribution as follows:

\begin{equation}
P_\theta = \frac{c n}{16 \pi r^2} {(\frac{r_+}{r})}^n (1-\theta) +
\frac{c}{16 \pi r_h} c^{\frac{2}{1-n}}
{(1-n)}^{\frac{n+1}{n-1}}{(1-\frac{2M_0}{r_+})}^{\frac{n+1}{n-1}}\delta
\end{equation}

For large values of $n$ the tangential pressure will tend to:

\begin{equation}
\lim_{n \rightarrow \infty}{P_\theta} = \frac{c n}{16 \pi r^2}
{(\frac{r_+}{r})}^n (1-\theta) + \frac{c(1-n)}{16 \pi r_+}
(1+\frac{r_0}{r_+})\delta
\end{equation}

with
$$ r_0 = - 2 M_0 $$

The "strong holographic selection principle" in combination with
the equation of state, $\rho + P_r = 0$, requires the interior
region to be free of tangential pressure. This leads to $n=0$. If
the membrane is to carry an energy content equal to the
gravitating mass of the compactar, and if $r_0 \ll r_+$, $c=1$
follows. So when the selection principle is applied to the
compactar solutions, we recover the holographic solution.

\section {\label{sec:linearP}Solutions with equation of state $P_r = a \rho$ }

In this section we would like to somewhat relax the constraint on
the equation of state. The main consideration for choosing an
equation of state linear in $P_r$ and $\rho$ is, that a linear
relationship renders a great part of the problem solvable in terms
of elementary functions.

It might appear to the reader that a linear equation of state is
overly restrictive. On the other hand, for highly gravitating
systems with highly relativistic particle momenta we should expect
that the pressure becomes comparable to the mass-density.
Furthermore any length-, mass- or time-scales related to the
electro-weak or strong forces will not be able to exert any
noticable influence on the physics of high gravitational fields. The
relationship between mass-density and pressure(s) should only
depend on dimensional considerations. The "natural" dimensional
relation between mass-density and the pressure is a linear one: In
units $c=1$ both have the same dimension.

In this section I will show that - for given $\rho$ - it is
possible to derive any solution with an equation of state $P_r = a
\rho$ from the special solution with a vanishing radial pressure.
The full class of solutions with a linear equation of state
therefore only depends on the mass-density $\rho$, on the
integration constant $M_0$ (or rather $r_0$) and on the constant
of proportionality between pressure and mass-density, $a$.

The metric coefficients of the solution with zero radial pressure
($a=0$) will be denoted by $A$ and $B$. The metric coefficients of
the general solution ($a \neq 0$) will be denoted by $A^{(a)}$ and
$B^{(a)}$, respectively. Likewise the tangential pressure for the
general case is denoted by ${P_\theta}^{(a)}$ and the tangential
pressure for the special case ($a=0$) by $P_\theta$.

$A$ does not depend on the radial pressure. Thus the radial metric
coefficient $A$ is independent of the equation of state parameter
$a$: $A^{(a)} = A$. We therefore drop the superscript in $A$. Only
$B^{(a)}$ and ${P_\theta}^{(a)}$ need to be determined.

It is easy to derive $B^{(a)}$:

$$ A B^{(a)} = e^{-\int_r^\infty{8 \pi r A (\rho + P_r)dr}}= e^{-(1+a)\int_r^\infty{8 \pi r A \rho dr}}= (A B)^{1+a}$$

\begin{equation}
B^{(a)} = A^a B^{1+a}
\end{equation}

We find that $B^{(a)}$ can be expressed by simple powers of $A$
and $B$, which is the metric for a radial pressure of zero. For
$a=-1$ the known result $B^{(-1)}=1/A$ is recovered.

The tangential pressure ${P_\theta}^{(a)}$ can be derived from the
tangential pressure of the zero (radial )pressure case. The
calculation is best done using the continuity equation. The result
is:

\begin{equation} \label{eq:Pt_a}
{P_\theta}^{(a)} = (1+a) P_\theta + a\left((\rho + \frac{r
\rho'}{2}) + (1+a) 2 \pi r^2 \rho^2 A \right)
\end{equation}

Lets take a closer look at the case $a=-1$. Due to the factor
$1+a$ in (\ref{eq:Pt_a}) the above expression is highly simplified:

\begin{equation} \label{eq:Pt:a-1}
{P_\theta}^{(-1)} = -(\rho + \frac{r \rho'}{2}) = -\frac{(r^2
\rho)'}{2r}
\end{equation}

For $a = -1$ the tangential pressure can be derived exclusively
from the mass density $\rho$ and it's first derivative.

By setting $\rho =  \frac{c}{8 \pi r^2} {(\frac{r_+}{r})}^n
(1-\theta)$, the results found in sections
\ref{sec:rho^n:a=-1:blackhole} and \ref{sec:rho^n:a=-1:compactar}
for $a = -1$ are recovered:

\begin{equation}
{P_\theta}^{(-1)} = \frac{n c}{16 \pi r^2} {(\frac{r_+}{r})}^n (
1-\theta) + \frac{c}{16 \pi r_+} \delta
\end{equation}

However, equation (\ref{eq:Pt:a-1}) is more general than the
result found in sections \ref{sec:rho^n:a=-1:blackhole} and
\ref{sec:rho^n:a=-1:compactar}. The mass-density $\rho$ isn't
limited to a power-expansion in $r$. Therefore equation
(\ref{eq:Pt:a-1}) immediately tells us, that - besides the trivial
case $\rho = 0$ - only an inverse square law for the mass-density
leads to a zero tangential pressure inside the entire
source-region. We can also see, that whenever the
mass-distribution is discontinuous, the tangential pressure will
have a $\delta$-distribution at the position of the discontinuity.

If $a \neq -1$ the field equations are non-linear and the weighted
superposition principle of section \ref{sec:general} cannot be
applied. It is not possible to construct a general solution by
power-expanding $\rho$ in the radial coordinate value $r$.
Nevertheless the solutions with $\rho \propto r^m$ provide
valuable hints with respect to the properties of more general
solutions.

\section{\label{sec:r^m}Solutions with $\rho \propto r^m$}

In this section I describe a general procedure for the derivation
of solutions with a mass-density following a power-law in $r$. The
mass-density is expressed as:

\begin{equation} \label{eq:rho:n:r^m}
\rho = \frac{c}{8 \pi r^2} \left(\frac{r_h}{r}\right)^n \,
(1-\theta)
\end{equation}

The radial metric coefficient for such a mass-density follows from
integration of equation (\ref{eq:r/A}). It is independent of the
equation of state:

\begin{equation} \label{eq:A:n:r^m}
A(r) = \frac{1}{1-\frac{r_h-r_0}{r} +
\frac{c}{n-1}((\frac{r_h}{r})^n - \frac{r_h}{r})} \, (1-\theta) +
\frac{1}{1-\frac{r_h-r_0}{r}} \, \theta
\end{equation}

The starting point of the integration has been chosen to be the
position of the boundary, $r_h$. The first term with $(1-\theta)$
describes the interior metric and corresponds to an integration in
the inward direction. The second exterior term (with $\theta$)
follows from integration in the outward direction. A positive
integration constant $r_0$ has been assumed at the boundary:

$$r_0 = \frac{r_h}{A(r_h)}$$

If the solution is to possess an event horizon, $r_0 = 0$.
Compactar solutions have $r_0 > 0$.

As long as the mass-density doesn't contain a
$\delta$-distribution the metric is continuous across the
boundary. By comparing the exterior metric with the Schwarzschild
metric we find:

$$r_h - r_0 = r_+ = 2M$$

The time coefficient of the metric can be determined from equation
(\ref{eq:AB'/AB}) as follows:

\begin{equation} \label{eq:lnAB}
\ln{AB}(r) = (1+a) 8 \pi \int_{r_h}^r{r A \rho \, dr} +
\ln{AB}(r_h)
\end{equation}

Due to the vanishing mass-density in the exterior space-time, the
exterior metric has $\ln{AB}(r) = \ln{AB}(r_h)= const$. Comparing
the exterior metric to the Schwarzschild-metric gives $AB(r\geq
r_h) =1$. Therefore the integration constant $\ln{AB}(r_h)$ must
be zero, if the metric is to be continuous across the boundary and
the exterior space-time is to be described by the Schwarzschild
metric.

For an arbitrary mass-density the integral in equation
(\ref{eq:lnAB}) for the interior metric cannot be expressed in
terms of elementary functions. For certain integer-powers a closed
form of the solution can be given. The most interesting cases are
discussed in the following sections.

\subsection {Solutions with $\rho = a P_r$ and a mass-density $\rho \propto 1/r^2$}

Let us assume an interior mass-density of the following form:

\begin{equation}
\rho(r) = \frac{c}{8 \pi r^2} (1-\theta)
\end{equation}

The integration of equation (\ref{eq:r/A}), starting out from the
boundary $r_h$ with $r_h / A(r_h) = r_0$ yields:

\begin{equation} \frac{r}{A} = \big(r_0 + (1-c) (r - r_h) \big)(1-\theta) +
\big(r - (r_h - r_0)\big)\theta
\end{equation}

From this $A$ can be determined:

\begin{equation} \label{eq:A:r^m}
A(r) = \frac{r}{r_0} \left({1- (1-c) \frac{r_h - r}{r_0}}\right)^{-1} (1-\theta) +
\left({1- \frac{r_h - r_0}{r}}\right)^{-1}\theta
\end{equation}

The exterior part of the solution (the term involving $\theta$) is
set equal to the Schwarzschild-metric with:

\begin{equation} r_+ = 2M = r_h - r_0
\end{equation}

$r_+$ is the gravitational radius and $M$ the total gravitating
mass.

The radial metric coefficient $A$ is continuous across the
boundary. $A$ reaches its maximum value at the boundary:

$$ A_{max} = A(r_h) = \frac{r_h}{r_0} = 1 + \frac{r_+}{r_0} = 1 + \frac{2M}{r_0}$$

Therefore the maximum value of $A$, at the position of the
boundary $r_h$, scales linearly with the mass of the compactar,
measured in Planck-units. A compactar/black hole of solar mass has
a gravitational radius of roughly 3 km. Therefore we have: $A(r_h)
\cong 3 km / l_{Pl} \approx 10^{38}$.

The time coefficient of the metric $B$ can be calculated by
integrating equation (\ref{eq:AB'/AB}). The integration is again
performed from the boundary, $r_h$:

$$ \ln{A(r) B(r)} - \ln{A(r_h) B(r_h)} = \int_{r_h}^r {8 \pi
r A(r) (\rho + P_r) dr} $$

The case $1+a = 0$ will not be discussed here. It has been
discussed in detail in the previous sections. For $c \neq 1$ we
get:

\begin{equation} \label{eq:AB}
\frac{AB(r)}{AB(r_h)} =  {(1 - (1-c) \frac{r_h -
r}{r_0})}^{-\frac{(1 + a) c}{c-1}}(1 - \theta) + \theta
\end{equation}

According to Birkhoff's theorem the exterior space-time ($\rho =
0$) should be described by the Schwarzschild metric. Therefore
$AB(r_h)$ must be $1$.

$B$ is finally given by:

\begin{equation} \label{eq:B}
B(r) =  \frac{r_0}{r} {(1 - (1-c) \frac{r_h -
r}{r_0})}^{1-\frac{(1 + a) c}{c-1}}(1 - \theta) + (1 - \frac{r_h
- r_0 }{r})\theta
\end{equation}

At the boundary, $B$ is just the inverse of $A$. Therefore
$B(r_h)$ can become very small for large compactars. For a
compactar of stellar mass $B(r_h) \approx 10^{-38}$.

The tangential pressure is calculated via equation
(\ref{eq:Ptan:AB}):

\begin{equation} \label{eq:Pt:A}
P_{\theta} = \frac{(1+a)c}{32 \pi r^2}\left((1+ac)A(r)-1\right)
(1-\theta) - \frac{a c}{16 \pi r_h} \delta
\end{equation}

By expanding $A$ we get the final result:

\begin{equation} \label{eq:Pt}
P_{\theta} = \frac{(1+a)c}{32 \pi r^2}\left(-1 +
(1+ac)\frac{r}{r_0} \left({1- (1-c) \frac{r_h -
r}{r_0}}\right)^{-1}\right)(1-\theta)- \frac{a c}{16 \pi r_h}
\delta
\end{equation}

For $c=1$:

\begin{equation} \label{eq:A:c=1} A(r) = \frac{r}{r_0} (1-\theta) +
\frac{1}{1- \frac{r_h - r_0}{r}}\theta
\end{equation}

\begin{equation} \label{eq:B:c=1}
B(r) =  \frac{r_0}{r} {e}^{-(1+a)\frac{r_h - r}{r_0}}(1 - \theta)
+ (1 - \frac{r_h - r_0 }{r})\theta
\end{equation}

\begin{equation} \label{eq:Pt:c=1}
P_{\theta} = \frac{1+a}{32 \pi r^2}\left(-1 + (1+a)\frac{r}{r_0}
\right)(1-\theta)- \frac{a}{16 \pi r_h} \delta
\end{equation}

The pressures and mass-densities at the boundary $r_h$ are given
by:

\begin{equation} \label{eq:Rho:rh}
\rho(r_h) = \frac{c}{8 \pi {r_h}^2}
\end{equation}

\begin{equation} \label{eq:Pr:rh}
P_r(r_h) = a \rho(r_h) = \frac{ac}{8 \pi {r_h}^2}
\end{equation}

\begin{equation} \label{eq:Pt:rh}
P_{\theta}(r_h) = \frac{(1+a)c}{4}  \left(-1 +
(1+ac)\frac{r_h}{r_0} \right) \rho(r_h) - \frac{a c}{16 \pi r_h}
\delta
\end{equation}

Note that for $c \neq 1$ the possible values of $r_h$ are somewhat
restricted. The radial metric coefficient $A(r)$ and the time
coefficient $B(r)$ should remain positive and real-valued
throughout the whole physically meaningful interior region of the
black hole. This leads to the following inequality:

\begin{equation} \label{eq:inequality}
1- (1-c)\frac{r_h-r}{r_0} \geq 0
\end{equation}

The "physically meaningful interior region" shall be defined as
$r>r_0$.\footnote{According to the discussion in
\cite{petri/charge} is appears more appropriate to choose
$r>r_0/2$. A slightly different choice of the "physically
meaningful interior region" affects the results only
quantitatively. The reader can easily make the necessary
adjustments.} If the above inequality is to be satisfied by all
interior r-values in the range between $r_0 \leq r \leq r_h$ we
get the following inequality:

\begin{equation} \label{eq:rhmax}
\frac{r_h}{r_0} \leq  \frac{2-c}{1-c}
\end{equation}

or

\begin{equation} \label{eq:cmin}
c \geq 1- \frac{r_0}{r_h-r_0} = \frac{1-2 \frac{r_0}{r_h}}{1-
\frac{r_0}{r_h}}
\end{equation}

If the interior mass-density is given by an inverse square law in
$r$, large compactars with $r_h \gg r_0$ are only possible if $c
\rightarrow 1$. For any compactar of a given size $r_h$ there is a
minimum value of $c$, given by equation (\ref{eq:cmin}), which
very rapidly approaches 1 for large $r_h$.

We can interpret this as tentative evidence, that large compactars
should have an interior mass-density very close to the "natural"
mass density $\rho = 1 / (8 \pi r^2)$.

Compactar solutions with a mass-density $\rho \propto 1/r^2$ are
of particular interest. The holostar solution with $\rho = 1 / (8
\pi r^2)$ will be discussed in detail in \cite{petri/hol}. The
holostar solution, in contrast to the more general solutions, can
be generalized in a natural manner to encompass charged matter.
The charged holostar solution is discussed in \cite{petri/charge}.
The interesting thermodynamic properties are discussed in
\cite{petri/thermo}. Some interesting properties of the more
general inverse square law solutions will be discussed in a
forthcoming paper.

\subsection{Solutions with $P_r = a \rho$ and $\rho \propto 1/r^4$}

Such solutions correspond to $n=2$ in equation
(\ref{eq:rho:n:r^m}). The radial metric coefficient has already
been calculated at the beginning of chapter \ref{sec:r^m}.

For $\rho \propto 1/r^4$ solutions for all values of $a$ and
$r_0$ can be expressed in terms of elementary functions. The
integration of the time-coefficient of the metric gives:

\begin{equation} \label{eq:lnAB:n2}
\ln{AB} = \frac{1+a}{2}\left(\frac{\gamma}{\beta} \ln{\delta
\frac{ (\beta+\gamma)\frac{r_h}{r}-2}
{(\beta-\gamma)\frac{r_h}{r}+2}} - \ln{\frac{r_h}{r_0}(1-\gamma
\frac{r_h}{r} + c \frac{r_h^2}{r^2})}\right)
\end{equation}

with

\begin{equation} \label{eq:gamma:n2}
\gamma = 1 + c -\frac{r_0}{r_h}
\end{equation}

\begin{equation} \label{eq:beta:n2}
\beta^2 = \gamma^2 - 4 c
\end{equation}

\begin{equation} \label{eq:delta:n2}
\delta = \frac{\beta + (2-\gamma)}{\beta -(2-\gamma)}
\end{equation}

$B$ is calculated by exponentiating equation (\ref{eq:lnAB:n2}).

The tangential pressure $P_{\theta}$ can be calculated from the
metric by a simple differentiation. It suffices to calculate
$P_{\theta}$ for $a=0$ and to derive the tangential pressure for
$a \neq 0$ by the procedure described in section
\ref{sec:linearP}. The calculation is straightforward, and can be
done either by hand or with a symbolic math program.

\subsection{Other solutions}

It is not possible to discuss, or even present, the full solution
space to the general spherically symmetric problem in a single
paper. For an arbitrary mass-density $\rho$ and an arbitrary
equation of state $P_r(\rho) = 0$, equations (\ref{eq:r/A} ,
\ref{eq:AB'/AB}) have to be integrated numerically. For some
particular cases solutions in terms of elementary functions are
possible. For $n=-1$, i.e. $\rho \propto 1/r$, there exists a full
solution for all values of $c$ and $r_0$ for a linear equation of
state. For $n=3$ and $n=-2$ black hole solutions (with $r_0=0$)
can be expressed in a closed form. For $n = -3$ there is a
solution for $r_0 = 0$ and $c=1$.

\section{Discussion}

A new class of solutions to the field equations of general
relativity has been presented. An unexpected property of the new
solutions is the appearance of a localizable two-dimensional
membrane at the generally non-continuous boundary of the matter
distribution. The membrane has a zero mass-density, but
considerable surface tension/pressure, hinting at a "new" state of
matter in high gravitational fields. However, if and how the new
solutions can contribute to a better understanding of the physical
phenomena of the Small and the Large remains a question, which can
only be answered in full by future research.

A field of research which presents itself immediately is the the
generalization of the solutions discussed in this paper to the
rotating and / or charged case. The charged holostar solution is
discussed in \cite{petri/charge}. The search for a solution
describing a compact rotating body will be a challenging topic of
future research.

The holographic solution appears to have some potential to
be a good approximation to our physical world. Its geometric
properties are discussed in detail in \cite{petri/hol}. In a
parallel paper \cite{petri/thermo} it is shown, that the
entropy/area law for black holes and the Hawking temperature
follow as a direct consequence of the holographic interior metric
($g_{rr} = r/r_0$), combined with microscopic statistical
thermodynamics.

Even if the holographic solution turns out to be incompatible with
the properties of the real world, it might still be of some
interest in another respect: There are indications that pressure
effects cannot be neglected in a realistic description of
gravitational phenomena, neither in the gravitational collapse of
a star, nor on a cosmological scale. For neutron stars anisotropic
models are already on the table. The simple mathematics of the
holographic solution will enable us to study some of the
consequences of pressure-induced effects and therefore might
provide valuable insights with respect to more realistic
space-times including (anisotropic) pressure.

The new solutions presented in this paper are exact mathematical
solutions of the field equations of general relativity. Any
solution that cannot be ruled out by convincing physical arguments
has a reasonable chance to be actually realized by nature. Due to
the high energies involved in self gravitational phenomena, the
usual scientific approach to choose among different mathematical
possibilities by well controlled experiments will not be feasible,
probably not even possible. It appears that ultimatively we will
have to decide by observation alone, which of the physically
acceptable solutions - if any - have been selected by nature. Most
likely the study of collision processes of compact self
gravitating objects, possibly the study of accretion processes,
will eventually provide unambiguous answers. Observation of such
effects will require sophisticated space- and ground-based
equipment. As long as conclusive observational evidence is
lacking, theoretical research will fill have to fill in the gaps.


\begin{thebibliography}{10}

\bibitem{ABCK}
Abhay Ashtekar, John~C. Baez, and Kirill Krasnov.
\newblock Quantum geometry of isolated horizons and black hole entropy.
\newblock {\em Adv. Theor. Math. Phys.}, 4:1--94, 2001, gr-qc/0005126.

\bibitem{Bardeen/Carter/Hawking}
J.~M. Bardeen, B.~Carter, , and S.~W. Hawking.
\newblock The four laws of black hole mechanics.
\newblock {\em Communications in Mathematical Physics}, 31:161, 1973.

\bibitem{Bekenstein/72}
Jacob~D. Bekenstein.
\newblock Black holes and the second law.
\newblock {\em Lett. Nuovo Cim.}, 4:737--740, 1972.

\bibitem{Bekenstein/73}
Jacob~D. Bekenstein.
\newblock Black holes and entropy.
\newblock {\em Physical Review D}, 7:2333--2346, 1973.

\bibitem{Bekenstein/Mukhanov}
Jacob~D. Bekenstein and V.~F. Mukhanov.
\newblock Spectroscopy of the quantum black hole.
\newblock {\em Phys. Lett. B}, 360:7--12, 1995, gr-qc/9505012.

\bibitem{Burinskii}
Alexander Burinskii, Emilio Elizalde, Sergi~R. Hildebrandt, and
Giulio Magli.
\newblock Regular sources of the {K}err-{S}child class for rotating and
  nonrotating black hole solutions.
\newblock {\em Physical Review D}, 65:064039, 2002, gr-qc/0109085.

\bibitem{Davies}
P.~Davies.
\newblock {\em The New Physics}.
\newblock Cambridge University Press, Cambridge, 1989.

\bibitem{Dev/00}
Krsna Dev and Marcelo Gleiser.
\newblock Anisotropic stars: Exact solutions.
\newblock {\em Gen. Rel. Grav.}, 34:1793--1818, 2002, astro-ph/0012265.

\bibitem{Dymnikova}
Irina Dymnikova.
\newblock Variable cosmological term - geometry and physics.
\newblock 2000, gr-qc/0010016.

\bibitem{Einstein/Grossmann}
A.~Einstein and M.~Grossmann.
\newblock Entwurf einer verallgemeinerten relativit\"{a}tstheorie und einer
  theorie der gravitation.
\newblock {\em Zeitschrift f\"{u}r Mathematik und Physik}, 62:225, 1914.

\bibitem{Elizalde}
Emilio Elizalde and Sergi~R. Hildebrandt.
\newblock The family of regular interiors for non-rotating black holes with
  {T00 = T11}.
\newblock {\em Physical Review D}, 65:124024, 2002, gr-qc/0202102.

\bibitem{Flie/AR}
T.~Flie{\ss}bach.
\newblock {\em Allgemeine {R}elativit\"{a}tstheorie, 3. {A}uflage}.
\newblock Spektrum Akademischer Verlag GmbH, Heidelberg Berlin, 1998.

\bibitem{Giambo}
Roberto Giamb\`{o}.
\newblock Anisotropic generalizations of de {S}itter spacetime.
\newblock {\em Class.Quant.Grav.}, 19:4399--4404, 2002, gr-qc/0204076.

\bibitem{Hawking/75}
S.~W. Hawking.
\newblock Particle creation by black holes.
\newblock {\em Communications in Mathematical Physics}, 43:199, 1975.

\bibitem{Herrera}
L.~Herrera, A.~Di Prisco, J.~Ospino, and E.~Fuenmayor.
\newblock Conformally flat anisotropic spheres in general relativity.
\newblock {\em J. Math. Phys.}, 42:2129--2143, 2001, qr-qc/0102058.

\bibitem{Strominger/Horowitz}
Gary Horowitz and Andrew Strominger.
\newblock Counting states of near-extremal black holes.
\newblock {\em Physical Review Letters}, 77:2368--2371, 1996, hep-th/9602051.

\bibitem{PlanckSolid}
Kenji Hotta.
\newblock The information loss problem of black hole and the first order phase
  transition in string theory.
\newblock {\em Prog.Theor.Phys.}, 99:427--450, 1998, hep-th/9705100.

\bibitem{Mak}
M.~K. Mak and T.~Harko.
\newblock Anisotropic stars in general relativity.
\newblock {\em Proc. Roy. Soc. Lond. A}, 459:393--408, 2003, gr-qc/0110103.

\bibitem{Maldacena/Strominger}
Juan Maldacena and Andrew Strominger.
\newblock Statistical entropy of four-dimensional extremal black holes.
\newblock {\em Physical Review Letters}, 77:428--429, 1996, hep-th/9603060.

\bibitem{Mazur/Mottola}
Pawel~O. Mazur and Emil Mottola.
\newblock Gravitational condensate stars: An alternative to black holes.
\newblock 2002, gr-qc/0109035.

\bibitem{Peacock}
P.~A. Peacock.
\newblock Cosmology and particle physics.
\newblock 2000.

\bibitem{petri/charge}
M.~Petri.
\newblock Charged holostars.
\newblock 2003, gr-qc/0306068.

\bibitem{petri/hol}
M.~Petri.
\newblock The holographic solution - Why general relativity must be understood in terms of strings.
\newblock 2004, gr-qc/0405007.

\bibitem{petri/thermo}
M.~Petri.
\newblock Holostar thermodynamics.
\newblock 2003, gr-qc/0306067.

\bibitem{Rovelli/measurement}
Carlo Rovelli.
\newblock A generally covariant quantum field theory and a prediction on
  quantum measurements of geometry.
\newblock {\em Nucl. Phys. B}, 405:797--815, 1993.

\bibitem{Rovelli/Smolin}
Carlo Rovelli and Lee Smolin.
\newblock Discreteness of area and volume in quantum gravity.
\newblock {\em Nucl. Phys. B}, 442:593--622, 1995, gr-qc/9411005.

\bibitem{Rovelli/primer}
Carlo Rovelli and Peush Upadhya.
\newblock Loop quantum gravity and quanta of space: a primer.
\newblock 1998, gr-qc/9806079.

\bibitem{Salgado}
Marcelo Salgado.
\newblock A simple theorem to generate exact black hole solutions.
\newblock {\em APS/123-QED, submitted to Class. Quant. Grav.}, 2003,
  gr-qc/0304010.

\bibitem{Smolin/advances}
L.~Smolin.
\newblock Recent developments in nonperturbative quantum gravity.
\newblock 2002, hep-th/9202022.

\bibitem{Smolin/measurement}
Lee Smolin.
\newblock Finite, diffeomorphism invariant observables in quantum gravity.
\newblock {\em Physical Review D}, 49:4028--4040, 1994, gr-qc/9302011.

\bibitem{Strominger/Vafa}
Andrew Strominger and C.~Vafa.
\newblock Microscopic origin of the bekenstein-hawking entropy.
\newblock {\em Physics Letters B}, 379:99--104, 1996, hep-th/9601029.

\bibitem{Susskind/hol}
L.~Susskind.
\newblock The world as a hologram.
\newblock {\em Journal of Mathematical Physics}, 36:6377, 1995, hep-th/9409089.

\bibitem{Hooft/hol}
G.~t'Hooft.
\newblock Dimensional reduction in quantum gravity.
\newblock 1993, gr-qc/9310026.

\bibitem{Thorne/mem}
Kip~S. Thorne, R.~H. Price, and D.~A. Macdonald.
\newblock {\em Black Holes: The Membrane Paradigm}.
\newblock Yale University Press, New Haven, Conneticut, 1986.

\bibitem{Weinberg/GR}
Steven Weinberg.
\newblock {\em Gravitation and Cosmology - Principles and Applications of the
  General Theory of Relativity}.
\newblock John Wiley and Sons, Inc., New York, 1972.

\bibitem{Zel'dovich/72}
Y.~B. Zel'dovich and A.A.Starobinsky.
\newblock Particle production and vacuum polarization in an anisotropic
  gravitational field.
\newblock {\em Soviet Physics - JETP}, 34:1159, 1972.

\end{thebibliography}

\newpage
\appendix

\section{The geodesic equations of motion for a spherically symmetric system}

If one studies the properties of a solution to the field
equations, it is often helpful to analyze the geodesic motion of
particles within the space-time. Note however, that in space-times
with significant pressure pure geodesic motion is not possible.

The purpose of this chapter is to express the geodesic equations
of motion for a general spherically symmetric system in a "more
geometric" form that will prove useful for future discussions.

In any spherically symmetric space-time the equations of geodesic
motion can be expressed in terms of the metric-coefficients $A$
and $B$ (see for example \cite[p. 135]{Flie/AR}):

\begin{equation} \label{eq:motion:r}
\left(\frac{dr}{d\tau}\right)^2 +
\frac{1}{A}(\frac{l^2}{r^2}+\epsilon) - \frac{F^2}{AB}= 0
\end{equation}

\begin{equation} \label{eq:motion:l}
\frac{d\varphi}{d\tau} r^2 = l = const
\end{equation}

\begin{equation} \label{eq:motion:F}
\frac{dt}{d\tau} B = F = const
\end{equation}

$\epsilon$ is zero for a particle of zero rest-mass and $1$ for a
massive particle. $l$ is the constant angular momentum (per
particle mass) and $F$ is a positive constant related to the total
energy of the motion of a massive particle.

It is quite helpful to express the motion of the particles as
fraction of the local velocity of light at a particular radial
position. The local velocity of light, expressed in terms of the
($t, r, \theta, \varphi$) coordinate values, can be read off from
the metric. The local velocity of light in the radial direction
($c_r$) and in the tangential direction ($c_\perp$) is given by:

\begin{equation} \label{eq:c:r}
c_r = \sqrt{\frac{B}{A}}
\end{equation}

\begin{equation} \label{eq:c:t}
c_\perp = \sqrt{B}
\end{equation}

The argument $r$ in the above quantities has been omitted.

If we denote by  $\beta_r(r)$ the local radial velocity of the
particle, i.e. expressed as ratio to the local speed of light in
the radial direction, and by $\beta_\perp(r)$ the respective local
tangential velocity, we find:

\begin{equation} \label{eq:beta:r}
\beta_r(r) = \frac{dr}{dt} / c_r = \frac{dr}{dt} /
\sqrt{\frac{B}{A}}
\end{equation}

and

\begin{equation} \label{eq:beta:t}
\beta_{\perp}(r) = \frac{r d\varphi}{dt} / c_\perp = \frac{r
d\varphi}{dt} / \sqrt{B} = \frac{l}{F} \frac{\sqrt{B}}{r}
\end{equation}

With this notation the radial equation of motion
(\ref{eq:motion:r}) can be written as a sum of a kinetic and
potential energy term:

\begin{equation} \label{eq:motion:r:2}
\beta_r^2(r) + V_{eff}(r) = 1
\end{equation}

with

\begin{equation} \label{eq:Veff:r}
V_{eff}(r) = \frac{B(r)}{F^2}(\frac{l^2}{r^2} +
\epsilon)=\beta_\perp^2(r) + \epsilon \frac{B(r)}{F^2}
\end{equation}

From equation (\ref{eq:Veff:r}) one can see, that the effective
potential can be expressed as the sum of two terms. The first
term, $\beta_{\perp}^2(r)$, is the square of the local tangential
velocity. The second term is only relevant for particles of
non-zero rest-mass. It involves $F^2$ and $B(r)$.

At any turning point of the motion, $r_i$, the radial velocity is
zero, and therefore according to equation (\ref{eq:motion:r:2}),
$V_{eff}(r_i) = 1$. This allows us to express the constants of the
motion, $l$ and $F$, by two parameters whose geometric
interpretation is more evident: the radial position of the turning
point, $r_i$, and the local tangential velocity at the turning
point, $\beta_{\perp}^2(r_i) = \beta_i^2$:

\begin{equation} \label{eq:l/F}
\frac{l^2}{F^2} = \frac{r_i^2 }{B(r_i)}\beta_i^2
\end{equation}

and

\begin{equation} \label{eq:F2}
\frac{\epsilon}{F^2} = \frac{1-\beta_i^2}{B(r_i)} =
\frac{1}{\gamma_i^2 B(r_i)}
\end{equation}

For particles with zero rest mass (photons) only equation
(\ref{eq:l/F}) is relevant. Equation (\ref{eq:F2}) is trivially
fulfilled. For photons the local tangential velocity at the
turning point of the motion must be equal to the local speed of
light in the tangential direction, i.e. $\beta_i^2 = 1$. For
particles with non zero rest-mass ($m_0 \neq 0$) the factor $
1-\beta_i^2 = 1 / \gamma_i^2$ is nothing else than the squared
ratio of the particle's rest mass to its local total relativistic
energy at the turning point of the motion, $E_i$. Therefore
equation (\ref{eq:F2}) can be expressed as follows:

\begin{equation} \label{eq:F2:2}
F^2 = B(r_i) \gamma_i^2 = B(r_i) {\left( \frac{E_i}{m_0}
\right)}^2
\end{equation}

In a pressure-free, stationary space-time the local energy of a
particle, measured by observers at different coordinate positions
is related by the gravitational Doppler-shift factor,
$\sqrt{g_{00}(r_1)/g_{00}(r_2)}$. For an asymptotically flat space
time $g_{00}(r=\infty) = B(\infty) = 1$. Under these circumstances
the constant of the motion $F$ can be identified with the local
total energy of the particle at its turning point of the motion,
divided by its rest-mass, as measured by an observer at rest at
spatial infinity.

The constant of the motion $l$ can be expressed as follows for a
particle of non-zero rest mass,

\begin{equation} \label{eq:l2:2}
l^2 = r_i^2 \frac{\beta_i^2}{1- \beta_i^2} = r_i^2 \beta_i^2
\gamma_i^2 = r_i^2 \left(\frac{p_i}{m_0}\right)^2
\end{equation}

where $p_i$ is the relativistic momentum of the particle at the
turning point of the motion. The geometric interpretation of $l$
is quite evident: The constant angular momentum $J$ of the
particle is given by $J = l m_0 = p_i r_i$, i.e. is proportional
to the local linear momentum $p$ times the radial coordinate
distance from the center $r$, both taken at the turning point of
the motion, $r_i$.

For further reference it is useful to express the local radial and
tangential velocities of particles undergoing geodesic motion in
the following form, involving only the "geometric" constants of
the motion $r_i$, $\beta_i$ and the metric coefficient $B$:

\begin{equation} \label{eq:beta:t:2}
\beta_{\perp}^2(r) = \frac{B(r)}{B(r_i)}\frac{r_i^2}{r^2}
\beta_i^2
\end{equation}

\begin{equation} \label{eq:beta:r:2}
\beta_r^2(r) + V_{eff}(r) = 1
\end{equation}

\begin{equation} \label{eq:Veff:beta}
 V_{eff}(r) = \frac{B(r)}{B(r_i)}\left(1 -
\beta_i^2(1-\frac{r_i^2}{r^2})\right)
\end{equation}

\end{document}